\documentclass[a4paper,11pt ]{article}
\usepackage{mathrsfs}
\usepackage{setspace}
\usepackage{graphics}
\usepackage{amsmath}
\usepackage{amssymb}
\usepackage{bm}
\usepackage{amsfonts}
\usepackage{ntheorem}
\usepackage{feynmf}
\usepackage{extarrows}
\usepackage{slashed}
\usepackage{graphicx}
\usepackage{indentfirst}
\usepackage{footmisc}
\usepackage{verbatim}
\begin{document}
\title{\bfseries{Moving stable solitons in  Galileon Theory}}
\author{\bfseries{Ali Masoumi\footnote{ali@phys.columbia.edu}, Xiao Xiao\footnote{xx2146@columbia.edu}}}

\maketitle
\begin{itshape}Physics Department and ISCAP,
Columbia University, New York, NY 10027, USA\end{itshape}
\begin{abstract}
Despite the  no-go theorem \cite{Derrick} which rules out  static stable solitons in Galileon theory, we propose a family of solitons that evade the theorem by traveling at the speed of light. These domain-wall-like solitons are stable under small fluctuations---analysis of perturbation shows neither ghost-like nor tachyon-like instabilities, and perturbative collision of these solitons suggests that they pass through each other asymptotically, which maybe an indication of the integrability of the theory itself.
\end{abstract}
\section{Introduction}
It is well-known that the special features of the DGP model \cite{DGP}, which is a brane-world model aimed at modifying gravity in the infrared, are encoded in the dynamics of a scalar degree of freedom which has a Galilean symmetry---invariance under a shift of the field variable composed  of a constant and a  part linear  in the  spacetime coordinates i.e $\phi \rightarrow \phi + a_0 + a_\mu x^\mu $.\cite{Consistency}. The Galilean model also arises in  massive gravity theories where  terms with Galilean invariance appear in the decoupling limit of massive graviton dynamics.\cite{massive gravity}\cite{Galileon}

It is an interesting question whether there are  stable solitons in Galileon theory.  In single field models, one could hope to find static solutions that  stay in a local minima of the energy functional, and that , although they are not protected by topology, their classical stability is still guaranteed. However Endlich et al.\cite{Derrick} proved that in a generic single field Galileon model in any number of  dimensions, there are no  stable static solutions. Their proof rules out localized static lumps in 3+1 dimensional spacetime and, because the argument is valid also in 1+1 dimensional spacetime, it also rules out static domain walls. In this paper we show that in 1+1 dimensions there is a family of stable solutions that evade this theorem by traveling at the speed of light and then we study their collisions. There is a domain-wall counterpart for them in higher dimensions. The result shows that the Galileon model accommodates the existence of domain walls, but because they are moving at the speed of light, they never look like static solutions. In section 2 we present the solutions and study their stability. In section 3 we find their energies and in section 4 we study the perturbative collision of these solitons and finally in section 5 we answer the possibility of existence of the solutions in higher dimensions. 
\section{The Solutions and Their Stability}
The Galileon Lagrangian is:
\begin{equation} \label{Lagrangian}
\mathscr{L}=\partial_{\mu}\phi\partial^{\mu}\phi+\alpha\partial_{\mu}\phi\partial^{\mu}\phi\Box\phi + \ldots ~.
\end{equation}
Because the higher order Galileon terms vanish in 1+1 dimensions,  we only present here the cubic interaction. The higher order terms  can also be dropped for the domain-wall solitons in  the 3+1 dimensional case.

The equation of motion is:
\begin{equation}\label{EOM}
\partial_{\mu}\partial^{\mu}\phi+\alpha[\left(\Box\phi\right)^{2}-\partial_{\mu}\partial_{\nu}\phi\partial^{\mu}\partial^{\nu}\phi]=0~.
\end{equation}

Written in light-cone coordinate, with $u=t+x$ and $v=t-x$, this equation simplifies to
\begin{equation}\label{LightconeEOM}
\frac{\partial^{2}\phi}{\partial u\partial v}+2\alpha\left[\left(\frac{\partial^{2}\phi}{\partial u\partial v}\right)^{2}-\frac{\partial^{2}\phi}{\partial u^{2}}\frac{\partial^{2}\phi}{\partial v^{2}}\right]=0 ~.
\end{equation}
It is obvious that if a function $\phi$ only depends on one of the light-cone coordinates,  the equation is satisfied and  we get the striking fact that any function of  $x\pm t$  is a solution of \eqref{EOM}. Therefore we can take any localized lump as the initial condition and let it propagate at the speed of light. There is no rest frame in which the solution is static, therefore it does not  contradict  Derrick's theorem.

Another way of seeing the existence of such solutions is to recall the fact that the Galileon theory has an interesting target-space diffeomorphism \cite{Euler Hierarchy 2}\cite{Euler Hierarchy}. The statement is that, for any solution $\phi_0(x,t)$ of the Galileon equation,  an arbitrary smooth function $f\left(\phi_{0}(x,t)\right)$ is also a solution of the Galileon equation if we only include the allowed highest order interaction in certain dimension\cite{Euler Hierarchy 2}\cite{Euler Hierarchy}\cite{closed curve}.In two dimensions or for domain-walls in higher dimensions, the highest relevant non-linear term is  the cubic one. Then one can see why these solutions exist: since $\phi_0(x,t) = x\pm t$ satisfies the Galilean equation, therefore $f(x\pm t)$ is also a solution for an arbitrary function $f$.

The main point of Derrick's theorem for the Galileon is ruling out the existence of the stable static solutions.  Here we prove  the perturbative stability of our non-static solutions against exponential growth of wave-like perturbations. We also eliminate the possibility of ghosts, waves with real frequency but the wrong sign in the action, by constraining the overall sign of the action perturbation. This is the approach adopted by Nicolis et al.\cite{Consistency}\cite{Stability} in studying the stability of the DGP model. We apply the same approach here.

Taking any classical solution $\phi_{0}(x)$, perturbing around the solution and calculating the deviation of the action, we get
\begin{equation}\label{ActionVariation}
\delta S=\int d^{4}x Z^{\mu\nu}\partial_{\mu}\delta\phi\partial_{\nu}\delta\phi ~.
\end{equation}
with
\begin{equation}
Z^{\mu\nu}=\eta^{\mu\nu}+2\alpha\Box\phi_{0}\eta^{\mu\nu}-2\alpha\partial^{\mu}\partial^{\nu}\phi_{0} ~.
\end{equation}
Since  $\phi_{0}$ is an extremum of the action, the term  linear in $\delta\phi$ vanishes, leaving only terms quadratic  and higher order in the perturbation.

The stability is determined by the tensor $Z^{\mu\nu}$. If we look at \eqref{ActionVariation} as describing a free particle, $Z^{\mu\nu}$ must have Minkowskian signature $(+,-,-,-)$ to ensure that the particle has physical dynamics. Usually it is simpler to use the metric to turn $Z^{\mu\nu}$ into $Z^{\mu}_{~\nu}$. The statement is then the $Z^{\mu}_{~\nu}$ should be positive-definite. If all of the  eigenvalues are positive after diagonalizing  $Z^{\mu}_{~\nu}$, then the solution is stable.

Written as a matrix,
\begin{equation}
Z^{\mu}_{~\nu}=\delta^{\mu}_{~\nu}-2\alpha\partial^{\mu}\partial_{\nu}\phi_{0}=\left(
                                                                                \begin{array}{cccc}
                                                                                  1-2\alpha\ddot{\phi_{0}} & -2\alpha\dot{\phi}'_{0} & 0 & 0 \\
                                                                                  2\alpha\dot{\phi}'_{0} & 1+2\alpha\phi_{0}''  &0 & 0\\
                                                                                  0  & 0 &    1    & 0  \\
                                                                                  0  & 0 &    0    & 1
                                                                                \end{array}
                                                                              \right) ~.
\end{equation}

First let's look at  the right-moving  solutions, i.e.  $\phi_{0}=f(x-t)$.

  In this case $Z_\nu^\mu$ simplifies to:
\begin{equation}
Z^{\mu}_{~\nu}=\left(
                \begin{array}{cccc}
                  1-2\alpha f'' & 2\alpha f'' &  0&   0\\
                  -2\alpha f'' & 1+2\alpha f''&  0 & 0 \\
                  0  & 0 &    1    & 0  \\
                  0  & 0 &    0    & 1
                \end{array}
              \right) \,\,.
\end{equation}
The eigenvalue equation
\begin{equation}
	\det ( Z - \lambda I )  =  (\lambda-1)^4=0 ~.
\end{equation}
only admits 1 as an eigenvalue. The only condition  we need for these solitons to be stable is the non-negativity of the eigenvalues, which is satisfied. This prevents the exponential growth of perturbations. However, there is a subtlety that we should address. The matrix $Z_\mu^\nu$ is not diagonalizable and it can only be upper-triangularized, which can be a matter of concern. However, we can do the same analysis by perturbing the equation of motion  and study growth of the perturbations over time.
\begin{equation}
	\phi (x,t) = f(x-t) + \delta \phi(x,t)~.
\end{equation} 
Plugging this into \eqref{EOM} gives:
\begin{equation}
	\delta\ddot{\phi}(x,t) \big( 1 - 2 \alpha f'' \big) - 4 \alpha f'' \delta\dot{\phi}' (x,t)- \delta \phi''(x,t) (1 + 2 \alpha f'')=0~.
\end{equation}
Assuming a wave  solution:
\begin{equation}
	\delta \phi(x,t) = A e^{i(k x  - \omega t)}~.
\end{equation}
leads to:
\begin{equation}
	\omega = \frac{2 \alpha k f'' \pm  k }{1 - 2 \alpha f''}~.
\end{equation}
This $\omega$ does not have an imaginary part and, based on the above arguments, we believe these solutions are locally stable. Here we treated $f''$ as constant which means that the wavelength of perturbation is assumed to be small compared to the length scales of the variation of $\phi(x)$. The same argument applies for the left moving solutions. For more details refer to \cite{Stability}~.

\section{Energy considerations}

The next step should be calculating the energy  of these solitons. We can calculate the energy-momentum tensor of these objects, but before proceeding, we need to address subtleties regarding the higher order time  derivative  terms in the action. 
Expanding the  Galileon Lagrangian  in \eqref{Lagrangian} gives
\begin{equation}
	S=\int dx dt \,\, \left[\dot{\phi}^{2}-|\nabla\phi|^{2}+\alpha\left(\dot{\phi}^{2}\ddot{\phi}-\dot{\phi}^{2}\nabla^{2}\phi-\ddot{\phi}|\nabla\phi|^{2}+\nabla^{2}\phi|\nabla\phi|^{2}\right) \right]~.
\end{equation}
We assume that the solution is trivial at temporal and spatial infinities.
The term $\dot{\phi}^{2}\ddot{\phi}$ is a total time derivative and can be dropped.  The $\ddot{\phi} \left| \nabla \phi\right|^2$ term can be simplified 
\begin{align}
	& \ddot{\phi} \left| \nabla \phi \right|^2 = \frac{d}{dt}\left( \dot{\phi} \left| \nabla \phi \right|^2\right) -2 \dot{\phi} \left(\nabla\phi \cdot \nabla \dot{\phi}\right) = \frac{d}{dt}\left( \dot{\phi} \left| \nabla \phi \right|^2\right) - \nabla \phi \cdot \nabla \dot{\phi}^2 \nonumber \\& \qquad = \frac{d}{dt}\left( \dot{\phi} \left| \nabla \phi \right|^2\right) - \nabla \cdot \left( \dot{\phi}^2 \nabla \phi\right) + \dot{\phi}^2 \nabla^2 \phi~.
\end{align}
Dropping the total derivatives  simplifies the  action to
 \begin{equation}
 	S=\int dx dt \left[ \dot{\phi}^{2}-|\nabla\phi|^{2}-2\alpha\dot{\phi}^{2}\nabla^{2}\phi+\alpha\nabla^{2}\phi|\nabla\phi|^{2}\right] ~.
 \end{equation}
 The conjugate momentum $\Pi$ and the Hamiltonian are 
 \begin{align}
   &\Pi=\frac{\delta S}{\delta\dot{\phi}}=2(1-2\alpha\nabla^{2}\phi)\dot{\phi} \\
   &H=\int dx  \left[\phi\dot{\phi}-L\right]=\int dx \left[\frac{\Pi^{2}}{4(1-2\alpha\phi_{,xx})}+|\phi_{,x}|^{2}-\alpha \phi_{,xx}|\phi_{,x}|^{2}\right]~.
 \end{align}
 We use the right moving solution $\phi(x,t) = f(x-t)$ to compute the Hamiltonian and get
\begin{equation}
E[f]=\int dx f'^{2}(2-3\alpha f'') \,\,.
\end{equation}
The second term is a total derivative and can be dropped and the energy simplifies to
\begin{equation}
E[f]=2\int dx f'^{2}\,\,.
\end{equation}
This energy is positive and remarkably does not depend on the interaction terms. The energy is exactly the same as if there were no higher derivative interactions in the Lagrangian in \eqref{Lagrangian}. The steeper the lump is, the larger the energy. The lump  energy scales as the inverse square of its width.
We can obtain the same conclusion by looking at the energy momentum tensor
\begin{equation}
T_{\mu\nu}=\frac{\delta S}{\delta \eta^{\mu\nu}}=\partial_{\mu}\phi\partial_{\nu}\phi+\alpha\Box\phi\partial_{\mu}\phi\partial_{\nu}\phi+\alpha\partial_{\lambda}\phi\partial^{\lambda}\phi\partial_{\mu}\partial_{\nu}\phi~.
\end{equation}
For the right moving solution $f(x-t)$, we have the stress tensor:
\begin{equation}
T_{\mu\nu}=\left(
  \begin{array}{cc}
    f'^{2} & -f'^{2} \\
    -f'^{2} & f'^{2} \\
  \end{array}
\right)\,\,.
\end{equation}
Again we see that the energy momentum tensor is independent of the non-linear terms proportional to $\alpha$ and therefore satisfies all the energy conditions for free waves moving at speed of light. Despite the fact that in higher dimensions  more interaction terms survive in Galileon lagrangian,  they do not change this energy formula and it remains correct in higher dimensions.

\section{ Perturbative collision of two solitons.}
\begin{figure}[htbp] 

   \includegraphics[width=3.4in]{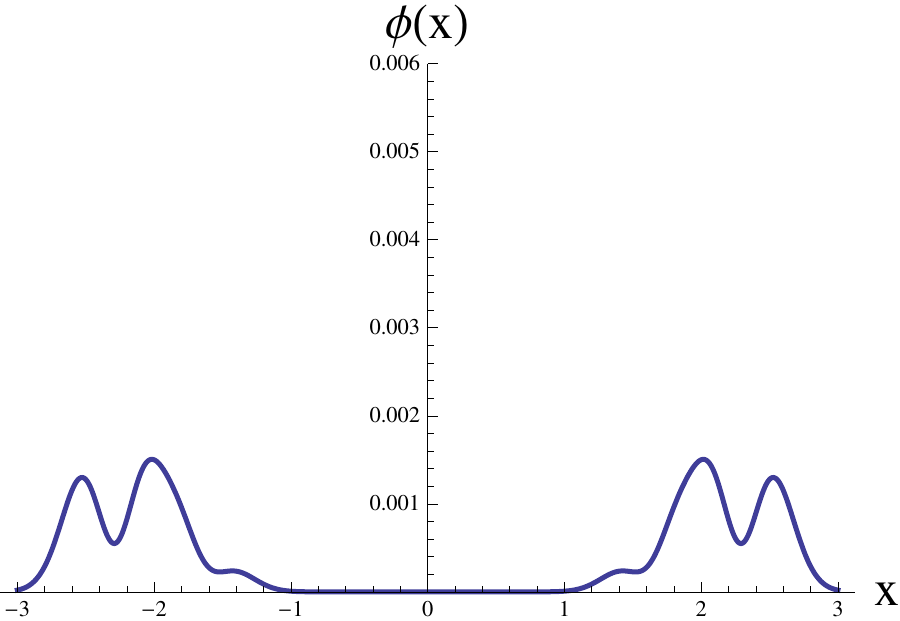}
   \includegraphics[width=3.4in]{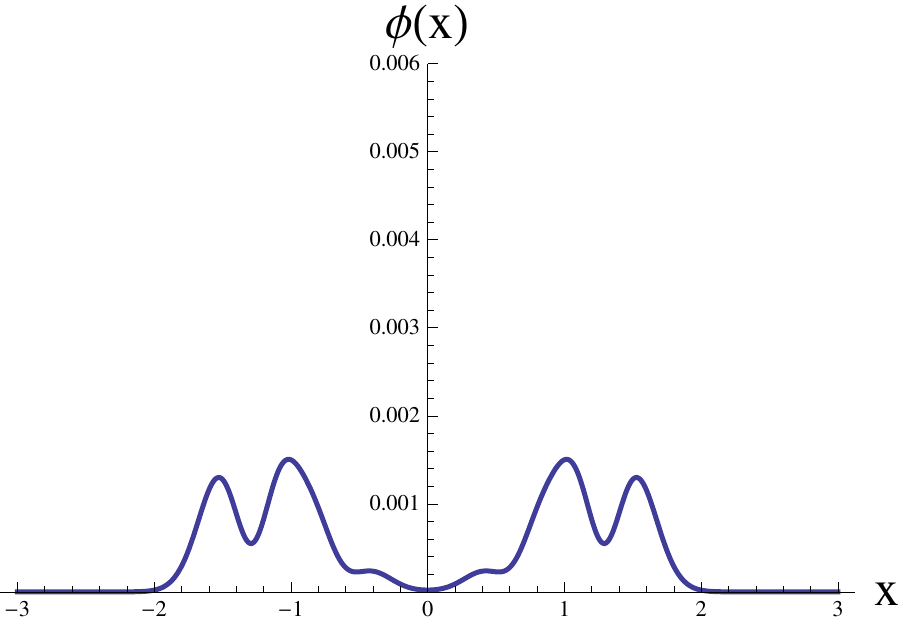}
   \includegraphics[width=3.4in]{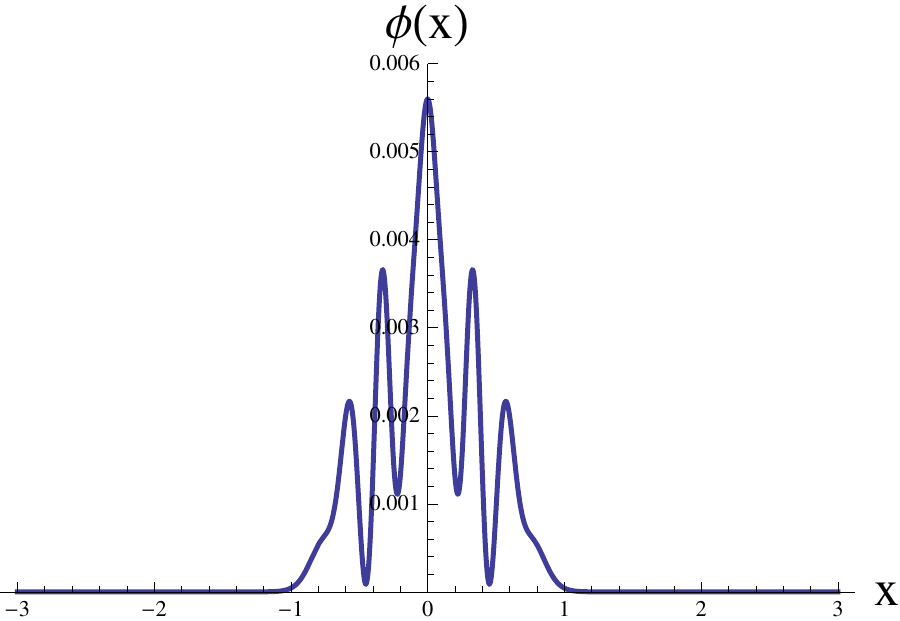}
   \includegraphics[width=3.4in]{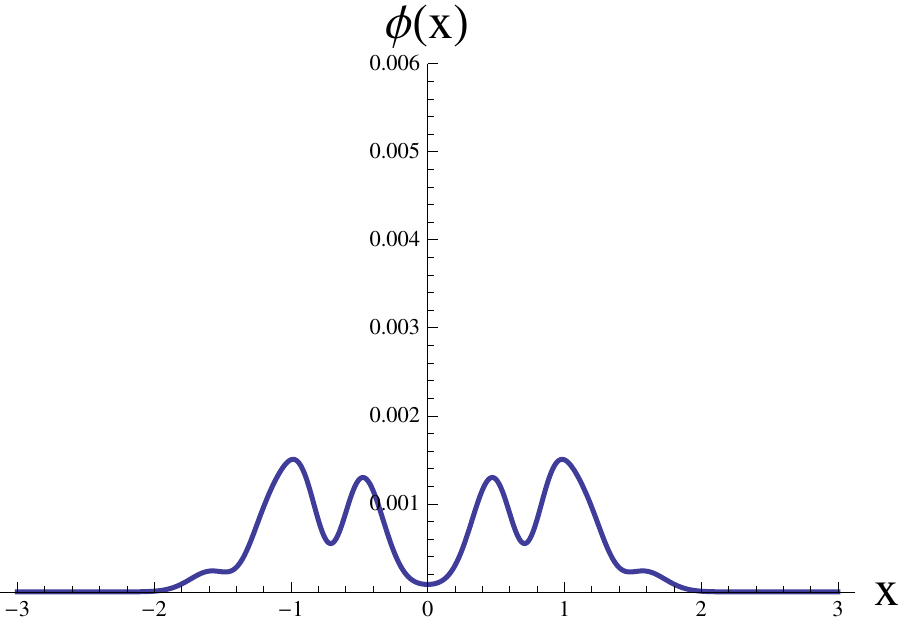}
   \includegraphics[width=3.4in]{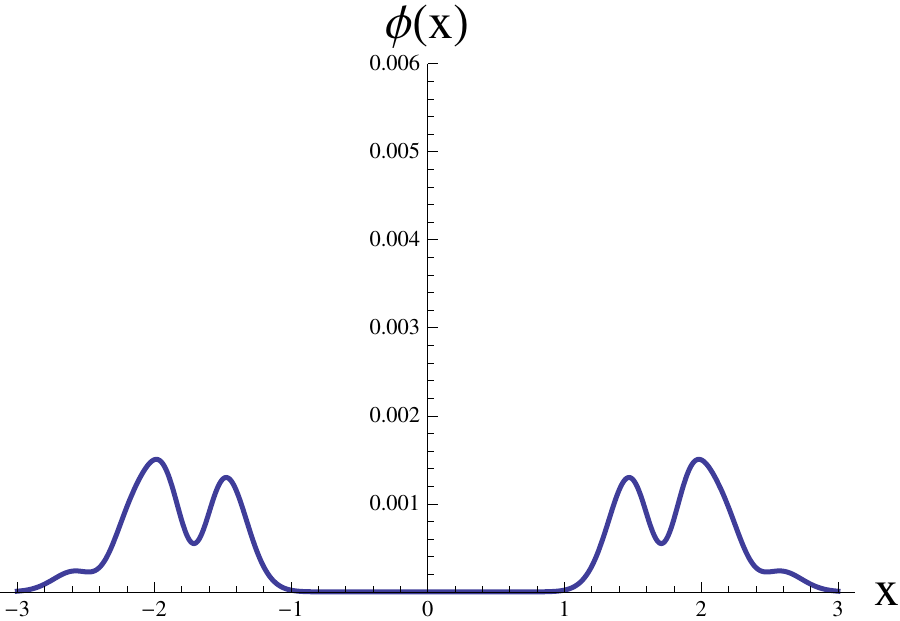}
   \caption{ Collision of the two solitons. The progression is from top left. As seen in these graphs, the two solitons retain their shape after they pass each other.}
   \label{Collision}
\end{figure}

One important aspect of studying solitons is their collision. If a right moving and a  left moving soliton collide, the result could be highly nontrivial. Let's assume that initially there are two such lumps moving towards each other without any interference term, so that
\begin{equation}
	\phi (x,t) = f(x+t) + g(x-t)\,\,.
\end{equation}
This wave-function does not satisfy the equations of motion. The full solution should be of the form:
\begin{equation} \label{SolutionAnsatz}
	\phi(x,t) = f(x+t) + g(x-t)  + h(x,t)	\,\,.
\end{equation}
We try to solve this equation in a limit $ |h(x,t)| \ll |\phi(x,t)|$ with boundary conditions $\lim_{t\rightarrow-\infty} |h(x,t)|= 0$.
The natural way to solve this equation is in the light-cone coordinates. Plugging \eqref{SolutionAnsatz} in  \eqref{EOM}, we obtain:
\begin{equation}
	 h_{,uv} ( 1 + 2 \alpha h_{,uv}) =2 \alpha (f'' + h_{,uu}) (g''+h_{,vv})=0\,\,.
\end{equation}
If $h$ and $\alpha$ are small, we can set up a perturbative expansion for this equation. Because $\alpha$ has dimensions of $[\text{Length}]^3$, we need to be careful about the meaning of its smallness.
 The only relevant length-scale in this problem is the width of the two lumps. In this approximation, the smallness of  $\alpha$ should be understood compared to the width of these lumps. If this criteria is not met, we cannot proceed much further. In this limit, we can iteratively integrate the equation of motion for $h$, assuming that it  vanished far in the past.\\
Expanding $h$ :
\begin{equation}
	h = h^{(1)}+ h^{(2)} + \ldots\,\,.
\end{equation}
The first order equation  :
\begin{align}
	&h^{(1)},_{u v} = 2  \alpha \, f''(u)\, g''(v) \,\,.
\end{align}
whose solution is:
\begin{align}
	&h^{(1)}(u,v) = 2 \alpha f'(u) g'(v) + 2 \alpha f'(u_0) g'(u_0) - 2 \alpha f'(u_0) g'(v) - 2 \alpha f'(u) g'(v_0)  \\
	& \qquad \qquad + h^1(u_0,v) + h^1(u,v_0) - h^1(u_0,v_0)\,\,.
\end{align}
where $u_0$ and $v_0$ are the initial points for the integration. Choosing these points in the far past, i.e. $u_0 , v_0 \rightarrow -\infty$,  and assuming that $f$ and $g$ are localized lumps, it is easy to see that all the terms on the right except for the first one vanish. Using the first order equation for $h$ ,  the second order equation is
\begin{align}
	&h^{(2)},_{u v} = 2 \alpha \, h^1_{,u u} g''(v) + 2 \alpha \, h^1_{,v  v} f''(u) \,\,.
\end{align}
which gives 
\begin{align}
	&h^{(2)}(u,v) = 2 \alpha^2 f''(u) g'^2(v) + 2 \alpha^2 f'^2(u) g''(v) + \text{Boundary terms}\,\,.
\end{align}
Using the fact that $h$ vanished at far past, the boundary terms should vanish. Similarly the 3rd order equation is:
\begin{align}
	&h^{(3)},_{u v} = 2 \alpha \big[ g'' h^2,_{u u} + f'' h^2,_{vv} + h^1,_{u u} h^1,_{v v} - 2 \alpha h^1,_{u v} f'' g''\big] \,\,.
\end{align}
leading to:
\begin{align}
	&h^{(3)}(u,v) = 4 \alpha^3 \big[  f'''(u)g'^3(v) + f'^3(u) g'''(v) + 6 f'(u) f''(u) g'(v) g''(v) \big] \\ \nonumber
	& \qquad \qquad - (2 \alpha)^3 \int_{-\infty}^u f''(u)^2 du \int_{-\infty}^v g''(v)^2 dv + \text{Boundary terms}\,\,.
\end{align}
It is important to notice that if the width of the lumps is large compared to $\alpha$, the higher-order terms in the perturbation, which contain higher derivatives, will be smaller . All the terms in the perturbation are interference terms between the right and left moving waves. Therefore, $h$ is negligible except for during the time that the two waves pass each other. When they get far enough apart, the interference term goes back to zero. Therefore in the limit of small $\alpha$ there is no dissipation and the two solitons retain their shapes. In Fig.\ref{Collision} we showed the collision obtained by this perturbative method before, during,  and after the collision of the two wave-packets of the form ($\alpha = 1$)
\begin{align}
& f(u) = A \big[u(u-\phi_1)(u-\phi_2)(u-\phi_3)(u-\phi_4)(u-\phi_5) +\phi_6 \big] \exp{[-\frac{u^2}{\text{width}^2}}]\\
& g(v) = A \big[v(v+\phi_1)(v+\phi_2)(v+\phi_3)(v+\phi_4)(v+\phi_5)-\phi_6 \big]\exp[{-\frac{v^2}{\text{width}^2}}]\,\,,
\end{align}
where
\begin{align*}
	&A=1 \qquad  \text{width} =0.3 \qquad \phi_1 = -0.3 \qquad \phi_2 = -0.1 \\
	&\phi_3=0.2\qquad \phi_4= 0.3 \,\,\, \qquad \phi_5=0.35 \qquad \phi_6=0.0015 \,\,.
\end{align*}
These graphs show that the two waves pass through each other perfectly and their shape and profile remains the same as before the collision after a long time. This can be an indication of the integrability of the Galilean Lagrangian as hinted by \cite{Euler Hierarchy 2}. 
\footnote{When this paper was in preparation, we were notified by K.Hinterbichler that the same solutions and their stability were independently discovered by J.Evslin\cite{closed curve}, who studied closed time-like curves in this model.}

\section{Conclusion}
With the above analysis, we have found a family of classical solitons in Galileon theory in (1+1) D. These solutions are  locally stable  and they have corresponding domain-wall-like structures in (3+1) dimensional spacetime. During collisions, these domain-walls would not change shape or dissipate, at least in the case that the width of the walls are large compared to the coupling constant $\alpha$. While there is a Derrick's theorem regarding the existence of stable static solutions in Galileon theory, we notice that stable ``moving solutions" are not necessarily ruled out and this work provides an example. Regarding the Derrick's theorem, another kind of solutions that  could not be ruled out are the gauged ones. Just as in the case of interactions via a potential,  gauging the theory \cite{Zhou:2011ix}\cite{Goon:2012mu} may violate the Derrick-type arguments and  allow the existence of stable static solutions. We also tried to find solutions analogue to a bubble moving at the speed of light in higher dimensions and could show the solutions are non-local and therefore not interesting.

\section{Acknowledgement}
We want to thank  Solomon Endlich, Lam Hui, Alberto Nicolis and Junpu Wang   for their comments and conversations. We especially want to thank Kurt Hinterbichler and Erick Weinberg  for reviewing this note.

\end{document}